\documentclass[prd,aps,showpacs,twocolumn,twoside]{revtex4-1}
\usepackage{amsfonts,amsmath,amssymb,bm,graphicx,subfigure,natbib}

\begin{document}

\title{Energy source for the magnetic field growth in magnetars driven by the electron-nucleon interaction}

\author{Maxim Dvornikov$^{a,b}$}
\email{maxdvo@izmiran.ru}

\author{Victor B. Semikoz$^{a}$}
\email{semikoz@yandex.ru}

\affiliation{$^{a}$Pushkov Institute of Terrestrial Magnetism, Ionosphere
and Radiowave Propagation of the Russian
Academy of Sciences (IZMIRAN), \\
142190 Troitsk, Moscow, Russia; \\
$^{b}$Physics Faculty, National Research Tomsk State University, \\
36 Lenin Avenue, 634050 Tomsk, Russia}

\date{\today}

\begin{abstract}
We study the magnetic field generation in a neutron star within the model based on the magnetic field instability in the nuclear matter owing to the electron-nucleon parity violating interaction. We suggest that the growing magnetic field takes the energy from thermal background fermions in the neutron star matter. The system of kinetic equations for the spectra of the magnetic helicity density and magnetic energy density as well as the chiral imbalance are solved numerically accounting for this energy source. We obtain that, for the initial conditions corresponding to a typical neutron star, the large scale magnetic field $\sim 10^{15}\thinspace\text{G}$ is generated during $(10^4-10^5)\thinspace\text{yr}$. We suggest that the proposed model describes strong magnetic fields observed in magnetars.
\end{abstract}




\maketitle

The most plausible explanation of radiation of soft gamma repeaters~\cite{Maz79} and anomalous X-ray pulsars~\cite{FahGre81} is the presence of strong magnetic fields $B\gtrsim 10^{15}\thinspace\text{G}$ in a neutron star (NS). Such highly magnetized NSs are called magnetars. Various models, explaining the origin of such strong astrophysical magnetic fields, were reviewed in Ref.~\cite{Fer15}. Nevertheless, the issue of the magnetic fields generation in magnetars still remains open.

Recently in Refs.~\cite{DvoSem15a,DvoSem15b} we proposed the new model for the generation of strong magnetic fields in magnetars based on the instability of magnetic fields in dense degenerate matter composed of nonrelativistic neutrons and ultrarelativistic electrons interacting by parity violating electroweak forces. The idea that electroweak interaction can induce the magnetic field instability was put forward first in Ref.~\cite{BoyRucSha12}. Within our model, basing on quite natural assumptions about the neutron star structure, we could describe the generation of large scale magnetic fields, with magnitudes predicted in magnetars, during time intervals comparable with magnetars ages.

Despite the plausibility of the model developed in Refs.~\cite{DvoSem15a,DvoSem15b}, it has a significant disadvantage. The instability of a magnetic field, proposed in Refs.~\cite{DvoSem15a,DvoSem15b}, is a necessary but not a sufficient condition for the magnetic field growth. To describe the magnetic field generation in magnetars one should indicate the source which feeds the magnetic field growth. This issue is addressed in the present work.

In this paper we further develop the model in Refs.~\cite{DvoSem15a,DvoSem15b}. We start with a brief description of the basic features of our model. Then we propose that magnetic fields can take the energy from the thermal motion of particles in the NS matter. We modify the kinetic equations, derived in Refs.~\cite{DvoSem15a,DvoSem15b}, to account for the magnetic field saturation, and numerically solve them. Finally, we discuss our results. In our work we use natural units in which $\hbar = c = k_\mathrm{B} = 1$.

Our model is based on the parity violating electroweak electron-nucleon
interaction (the $eN$ interaction). We shall take that the background
nuclear matter consists of neutrons and protons. This matter is supposed
to be unpolarized and nonmoving macroscopically. In Ref.~\cite{DvoSem15a} we derived 
the averaged effective Lagrangian of the $eN$ interaction in the Fermi approximation as
\begin{align}\label{eq:Hew}
  \mathcal{L}_{\mathrm{int}} = &
  - \bar{\psi}_{e}
  \gamma^0
  \left(
    V_\mathrm{L}P_\mathrm{L} + V_\mathrm{R}P_\mathrm{R}
  \right)
  \psi_{e},
  \notag
  \\
  V_\mathrm{L} = & \frac{G_{\mathrm{F}}}{\sqrt{2}}
  \left[
    n_{n} - n_{p}
    (1 - 4 \xi)
  \right] (1 - 2\xi),
  \notag
  \\
  V_\mathrm{R} = &
  - \frac{G_{\mathrm{F}}}{\sqrt{2}}
  \left[
    n_{n} - n_{p}
    (1 - 4 \xi)
  \right] 2\xi,
\end{align}
where $G_{\mathrm{F}}\approx1.17\times10^{-5}{\rm GeV}^{-2}$
is the Fermi constant, $n_{n,p}$ are the constant and uniform densities
of neutrons and protons, $\psi_{e}$ is the bispinor electron
wave function, $\xi = \sin^2\theta_\mathrm{W}\approx0.23$ is the Weinberg parameter, $P_\mathrm{L,R} = (1 \mp \gamma^5)/2$ are the chiral projectors, $\gamma^{5} = \mathrm{i} \gamma^{0}\gamma^{1}\gamma^{2}\gamma^{3}$, and $\gamma^{\mu} = (\gamma^{0},\bm{\gamma})$ are the Dirac matrices.

Now let us consider the interaction of ultrarelativistic electrons with background
matter, described by Eq.~(\ref{eq:Hew}), and an external magnetic
field $\mathbf{B}=(0,0,B)$. The total Lagrangian has the form, $\mathcal{L} = \mathcal{L}_{\mathrm{em}} + \mathcal{L}_{\mathrm{int}}$, where $\mathcal{L}_{\mathrm{em}} = \bar{\psi}_{e} \gamma^{\mu} \left( \mathrm{i}\partial_{\mu}+eA_{\mu} \right) \psi_{e}$ is the Lagrangian for the interaction of an ultrarelativistic electron with the electromagnetic field $A^{\mu}=\left(0,0,Bx,0\right)$, and $e>0$ is the absolute value of the electron charge.

The Dirac equation generated by $\mathcal{L}$ was solved in Refs.~\cite{DvoSem15a,DvoSem15b}. Using this solution, exactly accounting for both the matter interaction and the magnetic field, one can compute the induced electric current along the magneic field $J_z = - e \langle \bar{\psi}_{e} \gamma^3 \psi_e \rangle + \text{positron contribution}$, averaged using the Fermi-Dirac distribution. This current, which is additive to the ohmic current $\mathbf{J}_\mathrm{ohm}=\sigma_\mathrm{cond}\mathbf{E}$, where $\sigma_\mathrm{cond}$ is the matter conductivity and $\mathbf{E}$ is the electric field, turns out to be nonzero. If we restore the vector notations, one gets for the induced
electric current $\mathbf{J}=\Pi\mathbf{B}$. The parameter $\Pi$ reads
\begin{align}\label{eq:CSparam}
  \Pi = & \frac{2\alpha_{\mathrm{em}}}{\pi}
  \left(
    \mu_{5}+V_{5}
  \right),
  \notag
  \\
  \mu_{5} = & \frac{1}{2}
  \left(
    \mu_{\mathrm{R}}-\mu_{\mathrm{L}}
  \right),
  \notag
  \\
  V_{5} = & \frac{1}{2}
  \left(
    V_{\mathrm{L}}-V_{\mathrm{R}}
  \right) \approx
  \frac{G_{\mathrm{F}}}{2\sqrt{2}}n_{n},
\end{align}
where $\alpha_{\mathrm{em}} = e^2/4\pi \approx7.3\times10^{-3}$
is the fine structure constant. Note that, since we consider ultrarelativistic electrons, we can assume that right and left chiral projections of the electron-positron field behave independently and possess different chemical potentials: $\mu_{\mathrm{R}}$ and $\mu_{\mathrm{L}}$. To obtain Eq.~(\ref{eq:CSparam}) we assume that $n_{n}\gg n_{p}$
inside NS.

Using Eq.~\eqref{eq:CSparam}, in Ref.~\cite{DvoSem15b} we derived the system of kinetic equations for the spectra of the magnetic helicity densiy $h(k,t)$ and magnetic energy density $\rho_\mathrm{B}(k,t)$ as well as the chiral imbalance $\mu_5(t)$ in the form,
\begin{align}\label{general}
  \frac{\partial h(k,t)}{\partial t} = &
  -\frac{2k^2}{\sigma_\mathrm{cond}}h(k,t) +
  \left(
    \frac{4\Pi}{\sigma_\mathrm{cond}}
  \right)
  \rho_\mathrm{B}(k, t),
  \nonumber
  \\
  \frac{\partial \rho_\mathrm{B}(k,t)}{\partial t} = &
  -\frac{2k^2}{\sigma_\mathrm{cond}}\rho_\mathrm{B}(k,t) +
  \left(
    \frac{\Pi}{\sigma_\mathrm{cond}}
  \right)
  k^2 h(k, t),
  \notag
  \\
  \frac{\mathrm{d}\mu_5(t)}{\mathrm{d}t} = &
  \frac{\pi\alpha_\mathrm{em}}{\mu^2 \sigma_\mathrm{cond}}
  \int \mathrm{d} k
  \left[
    k^2h(k,t) - 2 \Pi \rho_\mathrm{B}(k,t)
  \right]
  \notag
  \\
  & -
  \Gamma_f\mu_5,
\end{align}
where $\mu$ is the chemical potential of electrons in NS, $\Gamma_f = 4 \alpha_\mathrm{em} m_e^2 /3\pi\sigma_\mathrm{cond}$ is the chirality flip rate in the electron-proton ($ep$) collisions, and $m_e$ is the electron mass. Note that the chirality flipping term in Eq.~\eqref{general} should contain $\mu_5$ since the equilibrium in the system of right and left electrons is achieved when $\mu_\mathrm{R}=\mu_\mathrm{L}$.

The total magnetic helicity $H$ and the magnetic field strength $B$ can be found on the basis of $h(k,t)$ and $\rho_\mathrm{B}(k,t)$ as
\begin{align}\label{hdef}
  H(t) = & \int \mathrm{d}^3 x (\mathbf{A} \cdot \mathbf{B}) =
  V \int h(k,t) \mathrm{d}k,
  \notag
  \\
  \frac{1}{2}B^2(t) = & \int \mathrm{d}k \rho_\mathrm{B}(k,t),
\end{align}
where $V$ is the normalization volume and the integration is over all the range of the wave number $k$ variation. It should be mentioned that in Eqs.~\eqref{general} and~\eqref{hdef} we assume the isotropic spectra.

In Ref.~\cite{DvoSem15b} we found that the model described by Eq.~\eqref{general} reveals the potential growth of the seed magnetic field $B_0 = 10^{12}\thinspace\text{G}$ up to $B \gtrsim 10^{17}\thinspace\text{G}$, i.e. the strengths predicted in magnetars. However, the energy source feeding the magnetic field growth was not specified in Ref.~\cite{DvoSem15b}. We demonstrate below that the magnetic field can take the energy from thermal motion of nucleons and electrons, which NS is composed of. For this purpose we shall calculate the temperature corrections to the energy density of degenerate fermions in NS as a possible source for the growth of the magnetic field. 

We shall start with the electron component of NS matter. Using the expansion of the integral in Ref.~\cite{LanLif80},
\begin{multline}\label{eq:intexp}
  \int_0^{\infty} \mathrm{d} \varepsilon
  \frac{f(\varepsilon)}{\exp[(\varepsilon - \mu)/T] + 1}
  \\
  =
  \int_0^{\mu}f(\varepsilon)\mathrm{d}\varepsilon +
  \frac{\pi^2}{6}T^2
  \left.
    \frac{\mathrm{d}f(\varepsilon)}{\mathrm{d}\varepsilon}
  \right|_{\varepsilon=\mu} +
  \mathcal{O}(T^4),
\end{multline}
one gets for the energy density
\begin{align}\label{endensel}
  \rho_e = &
  2 \int \frac{\mathrm{d}^3 p}{(2\pi)^3}
  \frac{p}{\exp[(p - \mu)/T] + 1} =
  \rho_{e0} + \delta \rho_{e},
  \notag
  \\
  \rho_{e0} = & \frac{\mu^4}{4\pi^2},
  \quad
  \delta \rho_{e} = \frac{\mu^2T^2}{2}, 
\end{align}
and the number density
\begin{align}\label{numdensel}
  n_e = &
  2 \int \frac{\mathrm{d}^3 p}{(2\pi)^3}
  \frac{1}{\exp[(p - \mu)/T] + 1} =
  n_{e0} + \delta n_{e},
  \notag
  \\
  n_{e0} = & \frac{\mu^3}{3\pi^2},
  \quad
  \delta n_{e} = \frac{T^2\mu}{3},
\end{align}
of degenerate ultrarelativistic electrons including temperature corrections. In Eqs.~\eqref{endensel} and~\eqref{numdensel} we keep only the leading terms in the temperature $T$. To derive Eqs.~\eqref{endensel} and~\eqref{numdensel} we neglect the magnetic fields correction to $\rho_e$ and $n_e$, studied in Ref.~\cite{Nun97}, since $eB \ll \mu^2$ for $B = (10^{12} - 10^{17})\thinspace\text{G}$ we consider here.

One can see in Eqs.~\eqref{endensel} and~\eqref{numdensel} that the mean energy of a thermal electron $\langle \varepsilon_e \rangle_T=\delta \rho_e/\delta n_e=3\mu/2$ exceeds the Fermi level $\mu$. The cooling of such electrons proceeds independently of the main contribution in degenerate electron gas with $0\leq \varepsilon_e\leq \mu$ since both the energy density of electrons and their number density are proportional to $T^2$. This cooling does not violate the Pauli principle for them either. That is why the decreasing of the temperature of such thermal electrons can feed the magnetic field growth.

Now let us consider degenerate nonrelativistic nucleons $N$, i.e. neutrons $N=n$ and protons $N=p$, as the energy source for the magnetic field growth. These particles have the Fermi energy $\mu_N=p_{F_N}^2/2M_N \gg T$, where $p_{F_N}$ is the nucleons Fermi momentum and $M_N$ is the nucleon mass. Analogously to Eqs.~\eqref{endensel} and~\eqref{numdensel}, as well as using Eq.~\eqref{eq:intexp}, we get the energy and number densities for degenerate nucleons, including thermal corrections, as
\begin{align}\label{endensN}
  \rho_N = & 
  2 \int \frac{\mathrm{d}^3 p}{(2\pi)^3}
  \frac{\varepsilon}{\exp[(\varepsilon - \mu_N)/T] + 1} =
  \rho_{N0} + \delta \rho_N,
  \notag
  \\
  \rho_{N0} = & \frac{p_{F_N}^5}{10\pi^2 M_N},
  \quad
  \delta \rho_N = \frac{T^2 M_N p_{F_N}}{4}, 
\end{align}
and
\begin{align}\label{numdensN}
  n_N = &
  2 \int \frac{\mathrm{d}^3 p}{(2\pi)^3}
  \frac{1}{\exp[(\varepsilon - \mu_N)/T] + 1} =
  n_{N0} + \delta n_N,
  \notag
  \\
  n_{N0} = &\frac{p_{F_N}^3}{3\pi^2},
  \quad
  \delta n_N = \frac{T^2 M_N^2}{6p_{F_N}}, 
\end{align}
where we account for the energy-momentum relation for a nonrelarivistic nucleon $\varepsilon = p^2 / 2 M_N$ and again keep only the leading terms in $T$.

Basing on Eqs.~\eqref{endensN} and~\eqref{numdensN}, one obtains the mean energy of thermal nucleons $\langle \varepsilon_N\rangle_T=\delta\rho_N/\delta n_N=3p_{F_N}^2/2M_N$, which is above the Fermi surface: $\langle \varepsilon_N\rangle_T > \mu_{N}$. Hence, like electrons, these nucleons can transfer their thermal energy to the magnetic field in their cooling without violation of the Pauli principle.

Summing up the thermal energy density corrections of electrons, protons, and nucleons, we can define the equipartition magnetic field strength $B_\mathrm{eq}$ as
\begin{align}\label{thermal}
  \frac{B_\mathrm{eq}^2}{2} = &
  \delta \rho_e + \delta \rho_p + \delta \rho_n
  \notag
  \\
  & =
  \left[
    \frac{M_n p_{F_n} + M_p p_{F_p}}{2} +\mu^2
  \right]
  \frac{T^2}{2}.
\end{align}
Accounting for $M_N \approx 940\thinspace\text{MeV}$, $p_{F_n} = (3 \pi^2 n_n)^{1/3} \approx 339\thinspace\text{MeV}$ for the NS density $n_n = 0.18\thinspace\text{fm}^{-3}$, and $p_{F_p} \approx \mu = 125\thinspace\text{MeV}$ for the electron density $n_e = 9 \times 10^{36}\thinspace\text{cm}^{-3}$,  one gets that the neutron contribution to $B_\mathrm{eq}$ is the greatest one. We can consider the quantity $\rho_\mathrm{T} = B_\mathrm{eq}^2/2$ in Eq.~\eqref{thermal} as the inexhaustible energy source requiring that 
$B_\mathrm{eq}^2\gg B^2$. Thus we do not violate the total energy conservation for the extended system which includes the background matter and the magnetic field. Since the values of $n_{n,e}$ are typical for NS we shall later use them in the numerical simulations.

In Refs.~\cite{DvoSem15a,DvoSem15b} we simulated magnetic fields in magnetars solving  Eq.~\eqref{general} and using $B_0 = 10^{12}\thinspace\text{G}$ as the initial magnetic field. Assuming $T=10^8\thinspace\text{K}$ and the above parameters of the NS matter, we get that $B_0^2 \ll B_\mathrm{eq}^2$. However, if $B = 10^{17}\thinspace\text{G}$, one obtains that $B^2 \gg B_\mathrm{eq}^2$. Thus strong magnetic fields, predicted in Refs.~\cite{DvoSem15a,DvoSem15b}, will influence the background matter in NS.

To avoid a back reaction on matter from such a strong magnetic field we should modify Eq.~\eqref{general}. As known from the solar dynamo theory~\cite{Cha14}, one can avoid the infinite growth of the magnetic field by quenching of the dynamo $\alpha$-parameter. Thus we can introduce the quenching of the parameter $\Pi$, given in Eq.~\eqref{eq:CSparam}, as
\begin{equation}\label{newPi}
  \Pi \to \frac{\Pi}{1+B^2/B_\mathrm{eq}^2},
\end{equation}
where $B^2$ and $B_\mathrm{eq}^2$ can be found in Eqs.~\eqref{hdef} and~\eqref{thermal}. Again referring to the solar dynamo theory, $B_\mathrm{eq}$ is equivalent to $B_{\odot} \sim 1\thinspace\text{kG}$, which is the magnetic field strength in a solar spot.
Now the excessive growth of the magnetic field is eliminated from our model since it is the parameter $\Pi$ which is responsible for the magnetic field instability.

To analyze the magnetic field generation in a magnetar on the basis of Eq.~\eqref{general} we should adopt an appropriate initial condition. The detailed discussion of the initial condition is provided in Ref.~\cite{DvoSem15b}. Here we just make a few comments on it.

We shall consider the evolution of a thermally relaxed NS at $t>t_0$, where $t_0 = 10^2\thinspace\text{yr}$. As obtained in Ref.~\cite{Yak11}, at $t_0<t\lesssim 10^6\thinspace\text{yr}$, NS cools down by the neutrino emission in modified Urca processes. The time dependence of the NS temperature obeys the differential equation~\cite{Pet92},
\begin{equation}\label{eq:law}
 \frac{\mathrm{d}T(t)}{\mathrm{d}t}= -\frac{T(t)}{(n_\mathrm{T}-2)t},
 \end{equation}
where the index $n_\mathrm{T} = 8$ for modified Urca processes. Using Eq.~\eqref{eq:law} and the results of Ref.~\cite{DvoSem15b}, one gets that the NS temperature and the NS conductivity will depend on time as $T^2 = T_0^2 F$ and $\sigma_\mathrm{cond} = \sigma_0/F$, where $F = (t/t_0)^{-1/3}$, $T_0 = 10^8\thinspace\text{K}$, and $\sigma_0 = 2.7\times10^8\thinspace\text{MeV}$ is given by the electron (or proton) density $n_e=n_p=9\times 10^{36}\thinspace{\rm cm}^{-3}$.

We shall study the generation of the magnetic field without specifying its direction, which can be random. Moreover we suggest that a seed magnetic field appears due to a turbulence which can be of a hydrodynamic origin.  In this case one can choose the initial Kolmogorov spectrum for the magnetic energy density $\rho_\mathrm{B}(k,t_0) = \mathcal{C} k^{-5/3}$~\cite{Dav04}. Here we correct the initial spectrum chosen in Ref.~\cite{DvoSem15b}. The constant $\mathcal{C}$ can be found from Eq.~\eqref{hdef} setting $B(t_0)=B_0 = 10^{12}\thinspace\text{G}$, which is a seed field typical for a young pulsar. The wave number runs in the interval $k_\mathrm{min} < k < k_\mathrm{max}$, where $k_\mathrm{min} = R_\mathrm{NS}^{-1} = 2\times10^{-11}\thinspace\text{eV}$, $R_\mathrm{NS} = 10\thinspace\text{km}$ is the NS radius, $k_\mathrm{max} = \Lambda_\mathrm{B}^{-1}$, and $\Lambda_\mathrm{B}$ is the free parameter specifying the scale of the magnetic field generated.

The initial spectrum of the helicity density can be chosen as $h(k,t_0)=2q\rho_\mathrm{B}(k,t_0)/k$, where the parameter $0 \leq q \leq 1$ defines the initial helicity. The case $q=0$ corresponds to the initially non-helical field and $q=1$ to the magnetic field with a maximal helicity. Therefore, besides magnetic fields we can also study the generation of the magnetic helicity in our model.

The initial value of the chiral imbalance can be taken as $\mu_5(t_0) = 1\thinspace\text{MeV}$. Note that $\mu_5(t_0)\neq 0$ is generated in direct Urca processes at the early stages of the NS evolution at $t<t_0$. The energy scale of these processes is governed by the mass difference between a neutron and a proton: $M_n - M_p \sim 1\thinspace\text{MeV}$. This fact substantiates our choice of $\mu_5(t_0)$.

At the first glance one can imagine that, for the chosen parameters, the contribution of the electroweak interactions $\sim V_5$ to Eq.~\eqref{general} is negligible compared to the electrodynamic contribution $\sim \mu_5$. As shown in Refs.~\cite{DvoSem15a,GraKapRed15}, almost any initial $\mu_5(t_0)\neq 0$ tends to zero very rapidly because of the high rate of $ep$ collisions, whereas $V_5$ is a steady source for the growth of the magnetic helicity and the magnetic energy density. Moreover, the electroweak term is not affected by $ep$ collisions since $V_5$ depends on the difference of the interaction potentials of left and right electrons with background matter, which are constant parameters of the model [see Eqs.~\eqref{eq:Hew} and~\eqref{eq:CSparam}], unlike $\mu_5$, which is a dynamic variable.

We also mention the recent Ref.~\cite{SigLei15}, where another steady source for the magnetic field instability, different from $V_5$, was used to explain strong magnetic fields in magnetars. It is based on the generation of the chiral imbalance in direct and modified Urca processes, $e_\mathrm{L}^- + p \to n + \nu_{e\mathrm{L}}$ and $e_\mathrm{L}^- + p +N \to n + \nu_{e\mathrm{L}} + N$, which are not in the equilibrium with inverse reactions. This situation can happen during $\sim 10\thinspace\text{s}$ after the onset of the supernova collapse outside the neutrinosphere.

Short scale, $\Lambda_\mathrm{B} \lesssim 1 \thinspace \text{cm}$, magnetic fields with the strength $B \lesssim 10^{14}\thinspace\text{G}$ were demonstrated in Ref.~\cite{SigLei15} to be generated in this situation. However, as shown in Ref.~\cite{MoiBis08}, short scale chaotic magnetic fields in a supernova explosion are subject to the reconnection with the typical time of several seconds. This time scale is comparable with the time interval for magnetic field generation in Ref.~\cite{SigLei15}. Thus, magnetic fields predicted in Ref.~\cite{SigLei15} will transform effectively into heat because of the magnetic reconnection.

\begin{figure*}
  \centering
  \subfigure[]
  {\label{1a}
  \includegraphics[scale=.26]{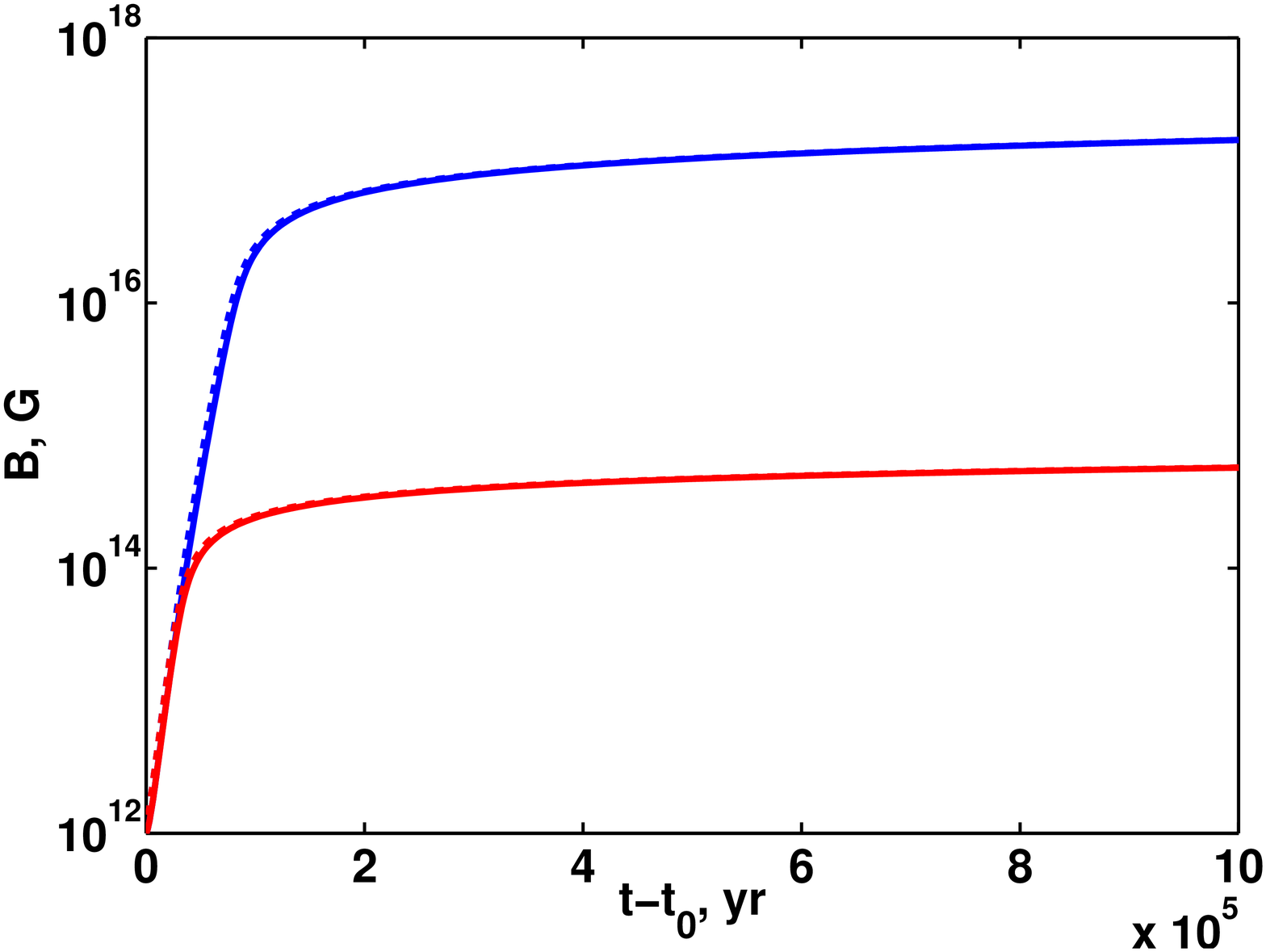}}
  \hskip-.7cm
  \subfigure[]
  {\label{1b}
  \includegraphics[scale=.26]{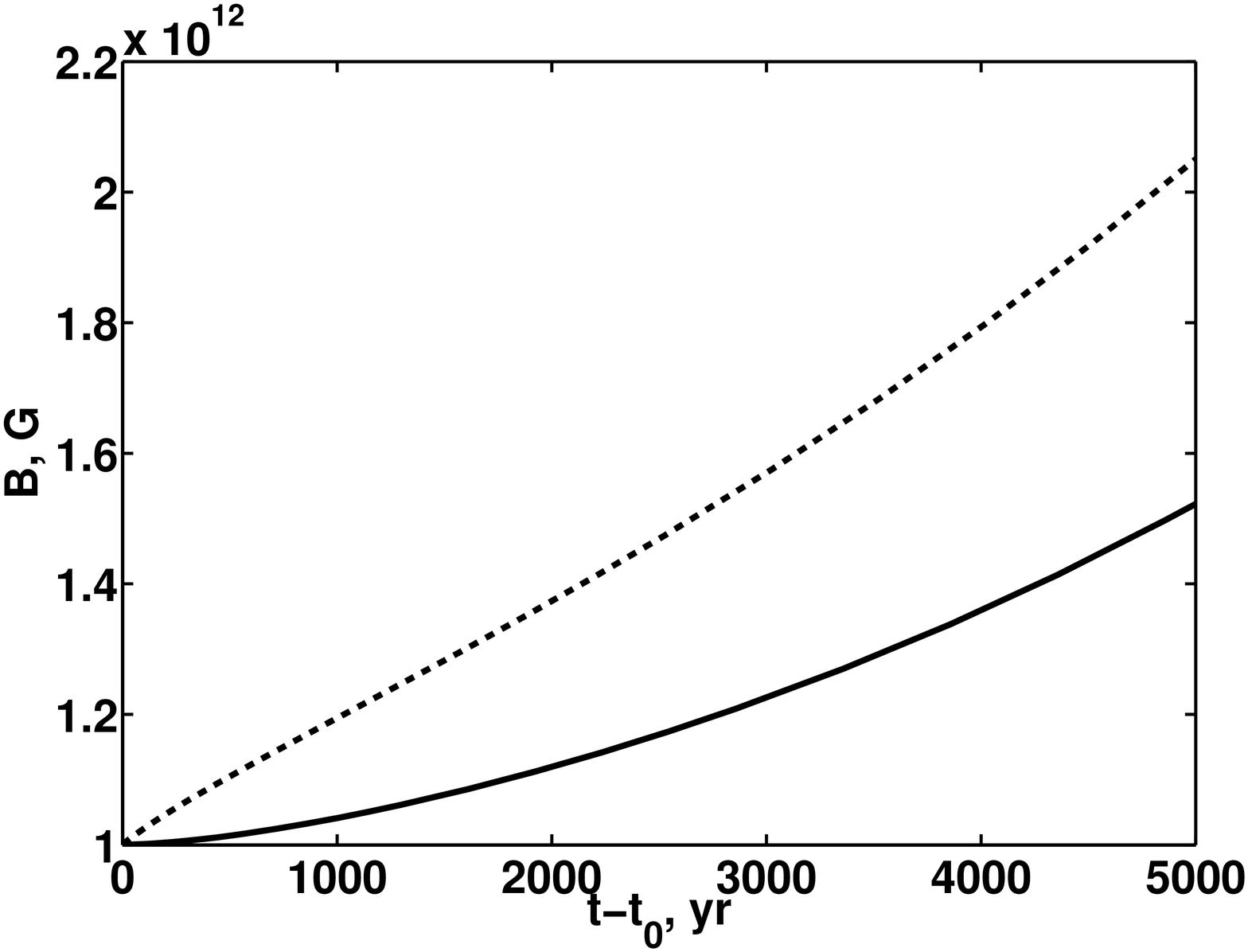}}
  \\
  \subfigure[]
  {\label{1c}
  \includegraphics[scale=.26]{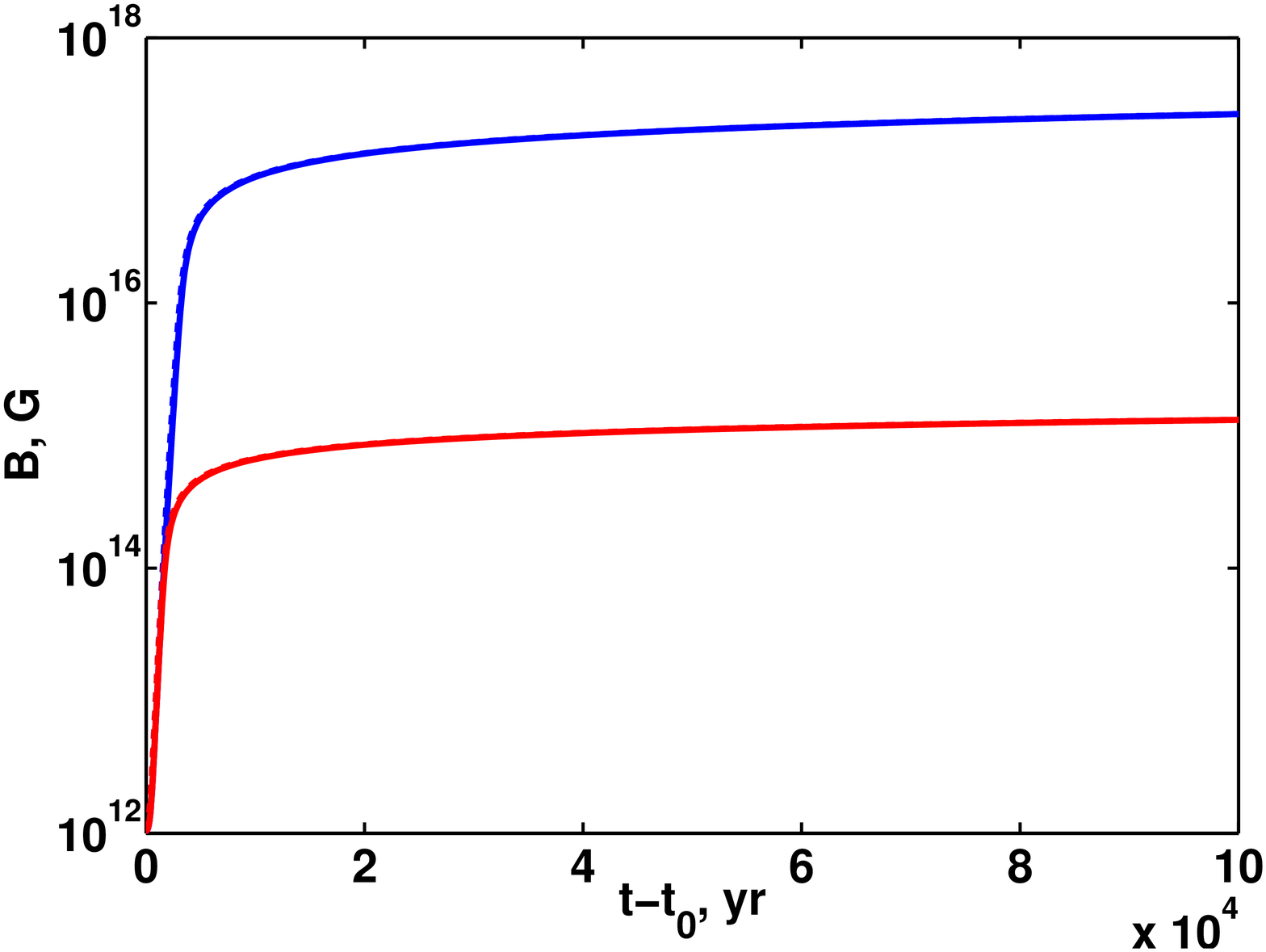}}
  \hskip-.7cm
  \subfigure[]
  {\label{1d}
  \includegraphics[scale=.26]{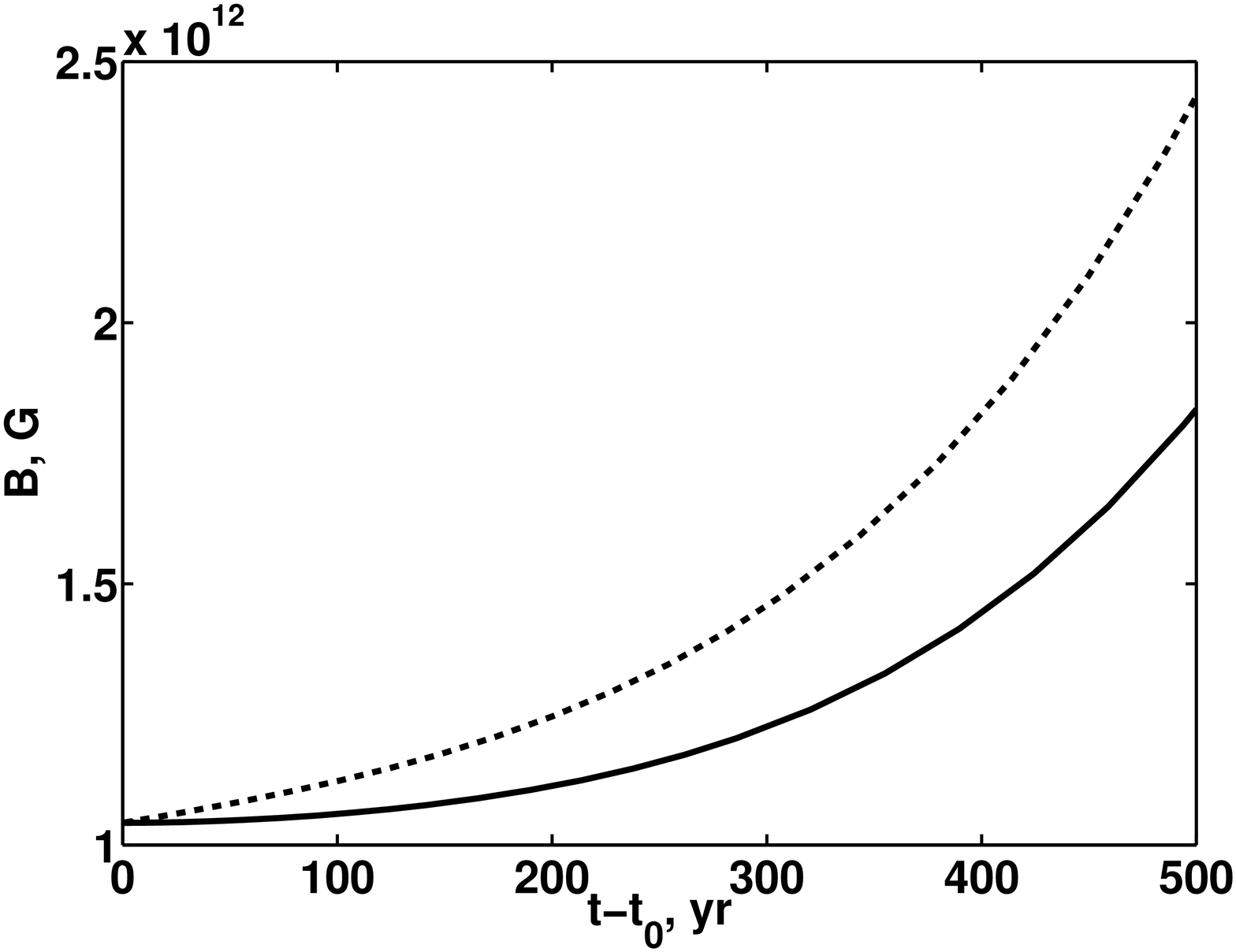}}  
    \caption{(color online).
    The growth of quenched and unquenched magnetic fields in
    magnetars versus $t-t_0$
    for different $k_\mathrm{max}$, corresponding to different length scales
    $\Lambda_\mathrm{B}$.
    Red and blue lines in panels~(a) and~(c) are the solutions
    of Eq.~\eqref{general} for quenched and unquenched $\Pi$ in   
    Eq.~\eqref{newPi} respectively.
    Solid lines correspond to initially nonhelical
    fields with $q=0$ and dashed lines to maximally helical fields with
    $q=1$.
    (a)~The magnetic field evolution for
    $k_\mathrm{max} = 2\times 10^{-10}\thinspace\text{eV}$ or
    $\Lambda_\mathrm{B}^{(\mathrm{min})} = 1\thinspace\text{km}$.
    (b) The behavior of the quenched magnetic field with the parameters as in 
    panel~(a) for shorter evolution time
    $t_0 < t < 5\times10^3\thinspace\text{yr}$.
    (c)~The magnetic field growth for
    $k_\mathrm{max} = 2\times 10^{-9}\thinspace\text{eV}$ or
    $\Lambda_\mathrm{B}^{(\mathrm{min})} = 100\thinspace\text{m}$.
   (d) The behavior of the quenched magnetic field with the parameters as in 
    panel~(c) for shorter evolution time
    $t_0 < t < 5\times10^2\thinspace\text{yr}$.
    \label{fig:B}}
\end{figure*}

Below we present the results of numerical solution of Eq.~\eqref{general} accounting for Eq.~\eqref{newPi} and the chosen initial conditions. In Fig.~\ref{fig:B} one can see the growth of magnetic fields of different length scales and initial helicities. We study the two main minimal scales: $\Lambda_\mathrm{B}^{(\mathrm{min})} = 1\thinspace\text{km}$ in Figs.~\ref{1a} and~\ref{1b} as well as $\Lambda_\mathrm{B}^{(\mathrm{min})} = 100\thinspace\text{m}$ in Figs.~\ref{1c} and~\ref{1d}. Thus we predict the generation of strong large scale magnetic fields.

To compare the behavior of magnetic fields in the present work with that in Ref.~\cite{DvoSem15b}, in Figs.~\ref{1a} and~\ref{1c} we also show the results of the numerical solution of Eq.~\eqref{general} without quenching in Eq.~\eqref{newPi}. One can see that unquenched magnetic fields, shown by blue lines, slow down the growth rate after $\sim 10^5\thinspace\text{yr}$ in Fig.~\ref{1a} and $\sim 10^4\thinspace\text{yr}$ in Fig.~\ref{1c}, but continue growing~\cite{DvoSem15b}. On the contrary, the quenched magnetic fields, shown by red lines, are saturated. For both $\Lambda_\mathrm{B}^{(\mathrm{min})}$ we start with $B_0 = 10^{12}\thinspace\text{G}$ and magnetic fields reach the saturated value $B_\mathrm{sat} \sim 10^{15}\thinspace\text{G}$. For example, in Fig.~\ref{1b}, $B_\mathrm{sat} \approx 1.1\times 10^{15}\thinspace\text{G}$. This $B_\mathrm{sat}$ is close to magnetic fields observed in magnetars~\cite{Mer15}.

Magnetic fields in Fig.~\ref{1a} grow up to $B_\mathrm{sat}$ for $t\gtrsim 10^5\thinspace\text{yr}$ and in Fig.~\ref{1c} for $t\gtrsim 10^4\thinspace\text{yr}$. These time intervals are comparable with the ages of young magnetars~\cite{Mer15}. Note that the smaller the scale of the magnetic field is, the faster this magnetic field grows and the stronger $B_\mathrm{sat}$ is. One gets from Eq.~\eqref{hdef} that $\rho_\mathrm{B} \sim k^2 A^2$, where $A$ is the typical vector potential. Hence, a bigger $k_\mathrm{max}$ corresponds to stronger $B_\mathrm{sat}$.

We also analyze the evolution of magnetic fields with different initial helicities. One can see in Figs.~\ref{1b} and~\ref{1d} that there is a difference in the behavior of magnetic fields for initially non-helical (solid lines) and maximally helical (dashed lines) fields for relatively small evolution times. At later times this difference is washed out; cf. Figs.~\ref{1a} and~\ref{1c}. It means that, in frames of our model, we can generate both strong magnetic fields and the magnetic helicity. 

Note that the behavior of quenched and unquenched magnetic fields is almost indistinguishable at small evolution times. Indeed, if $t \ll t_\mathrm{sat}$, where $t_\mathrm{sat} = (10^4 - 10^5)\thinspace\text{yr}$ is the saturation time depending on the scale of the magnetic field, then $B \ll B_\mathrm{eq}$ in Eq.~\eqref{thermal}. Thus in this time interval it is sufficient to consider the evolution of quenched magnetic fields, which is shown in Figs.~\ref{1b} and~\ref{1d}.

Comparing the results of Ref.~\cite{DvoSem15b} with the evolution of magnetic fields in Figs~\ref{1b} and~\ref{1d}, one can notice that in the present work  magnetic fields grow several times slower. This discrepancy can be attributed to the fact that now we use the correct initial Kolmogorov's spectrum $\rho_\mathrm{B}(k,t_0) = \mathcal{C} k^{\nu_\mathrm{B}}$ with
$\nu_\mathrm{B} = -5/3$ vs. $\nu_\mathrm{B} = 1/3$ in Ref.~\cite{DvoSem15b}. Indeed, since $\mathcal{C} \sim \nu_\mathrm{B}+1$, the greater $\nu_\mathrm{B}$ is, the faster $\rho_\mathrm{B}(k,t)$ will grow. 

\begin{figure*}
  \centering
  \subfigure[]
  {\label{2a}
  \includegraphics[scale=.26]{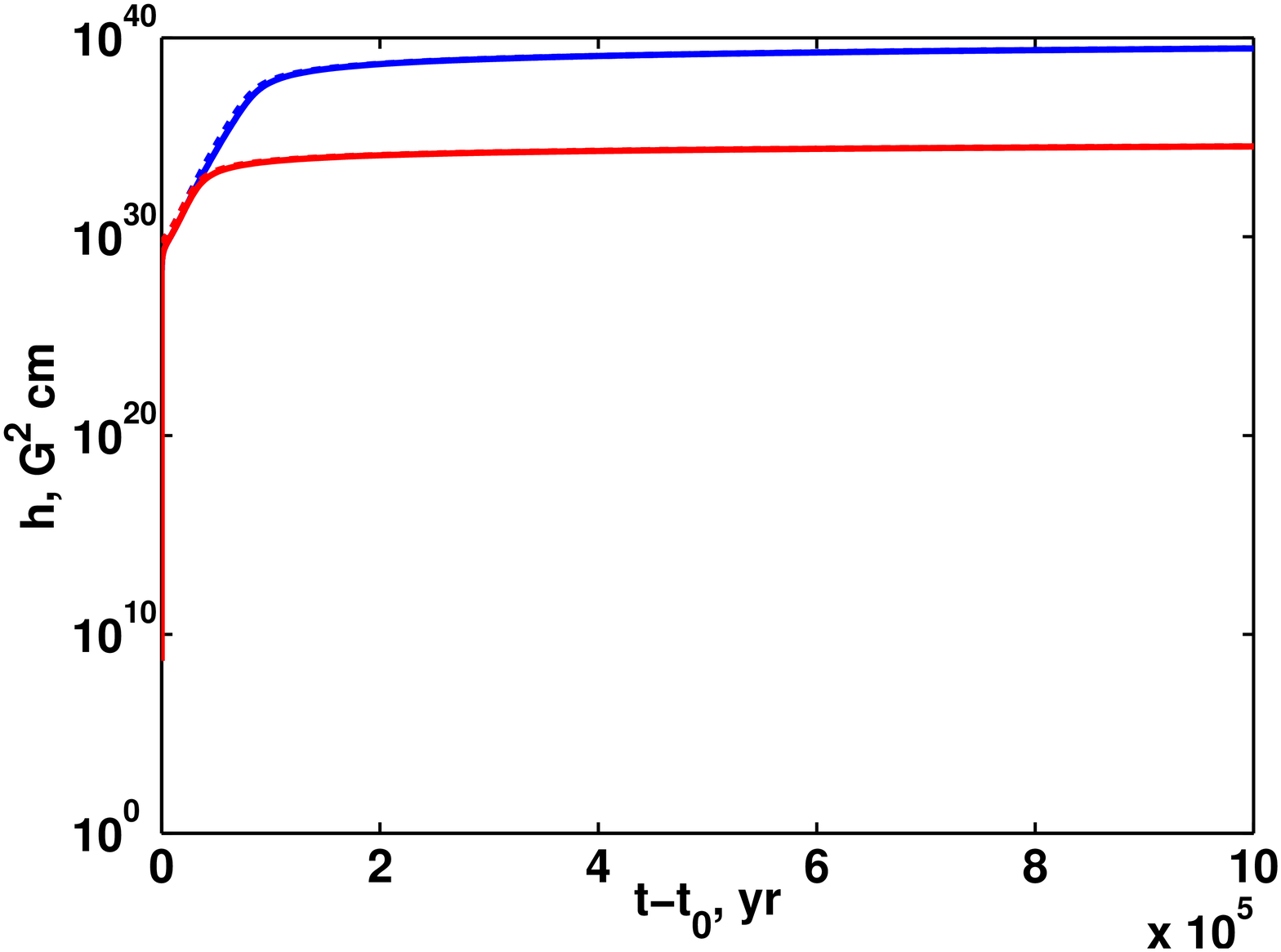}}
  \hskip-.7cm
  \subfigure[]
  {\label{2b}
  \includegraphics[scale=.26]{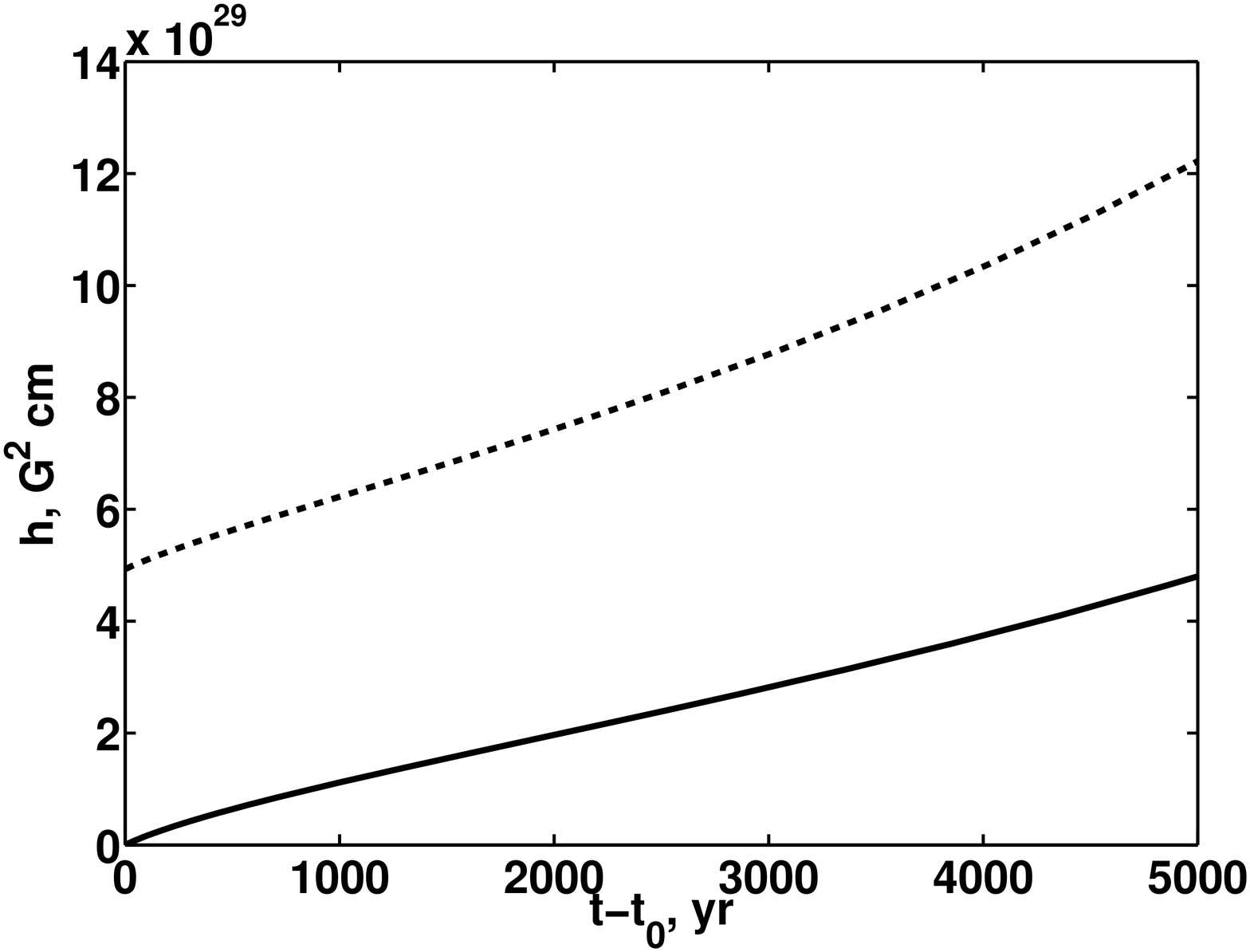}}
  \\
  \subfigure[]
  {\label{2c}
  \includegraphics[scale=.26]{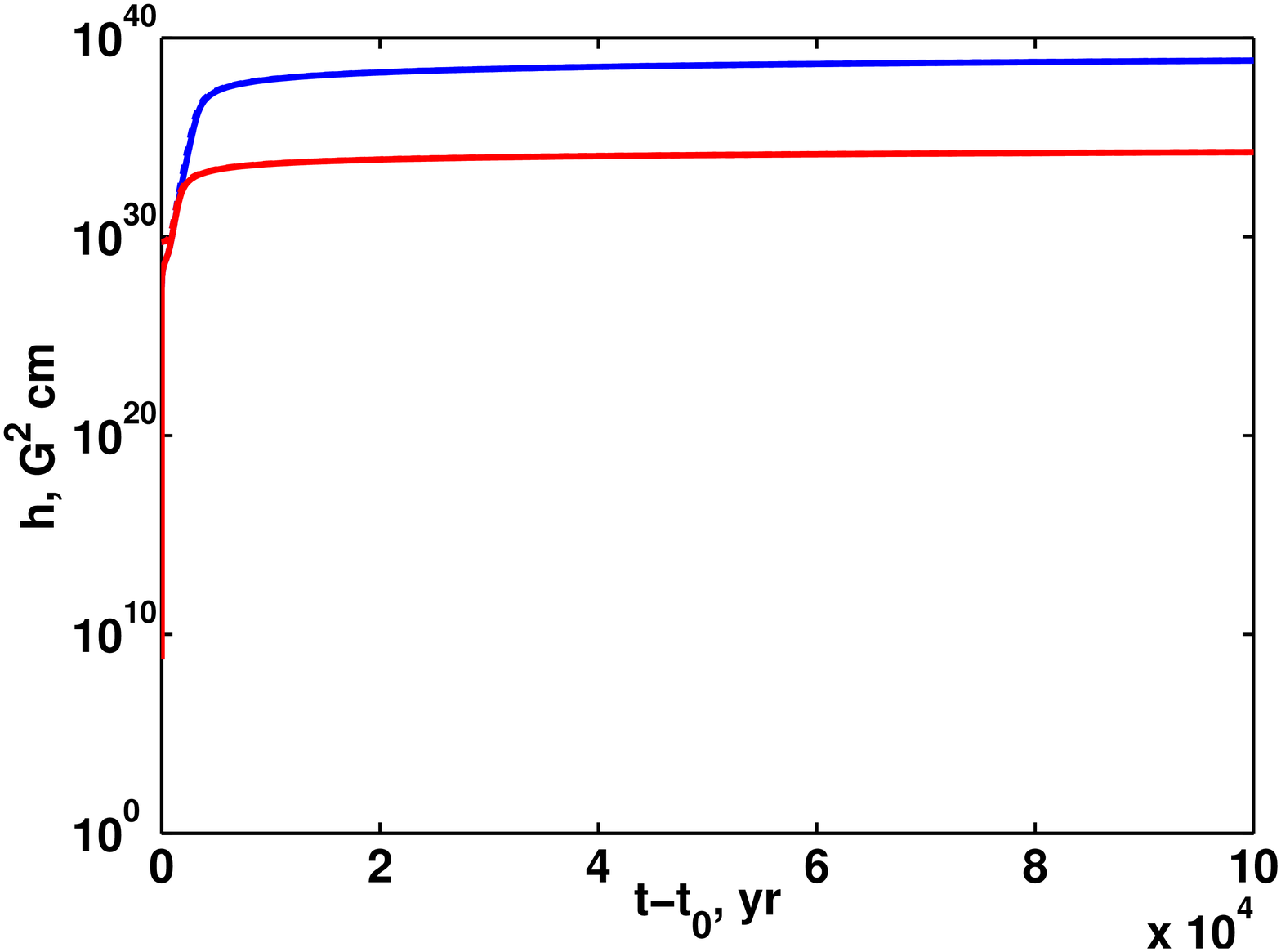}}
  \hskip-.7cm
  \subfigure[]
  {\label{2d}
  \includegraphics[scale=.26]{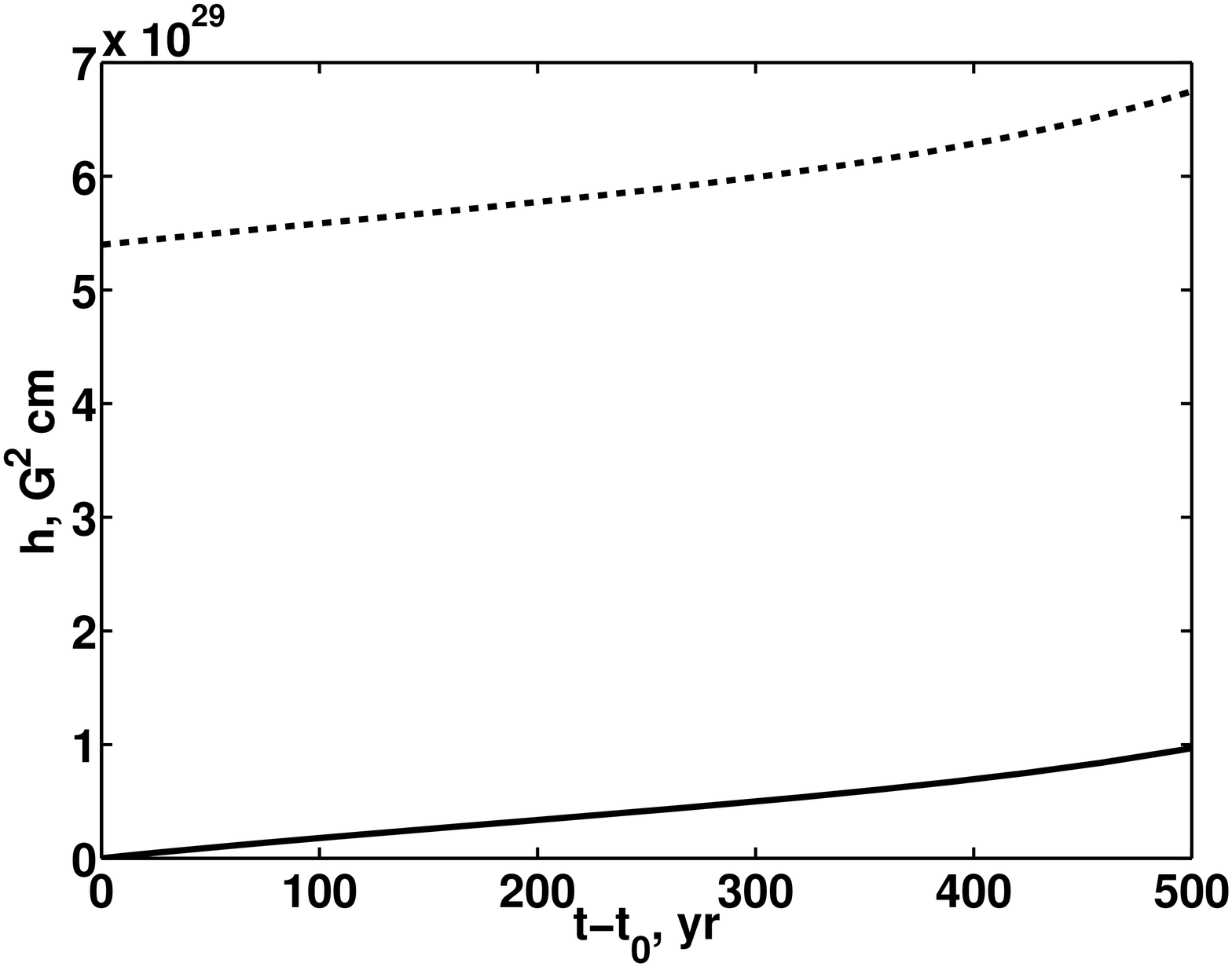}}  
    \caption{(color online).
    The evolution of the magnetic helicity in
    magnetars versus $t-t_0$ for quenched and unquenched parameter $\Pi$ 
    as well as
    for different $k_\mathrm{max}$, corresponding to different length scales
    $\Lambda_\mathrm{B}$.
    Red and blue lines in panels~(a) and~(c) are the solutions
    of Eq.~\eqref{general} for quenched and unquenched $\Pi$ in   
    Eq.~\eqref{newPi} respectively.
    Solid lines correspond to initially nonhelical
    fields with $q=0$ and dashed lines to maximally helical fields with
    $q=1$.
    (a)~The magnetic helicity density evolution for
    $k_\mathrm{max} = 2\times 10^{-10}\thinspace\text{eV}$ or
    $\Lambda_\mathrm{B}^{(\mathrm{min})} = 1\thinspace\text{km}$.
    (b) The behavior of the magnetic helicity density for the quenched $\Pi$
    with the parameters as in panel~(a) for shorter evolution time
    $t_0 < t < 5\times10^3\thinspace\text{yr}$.
    (c)~The magnetic helicity density growth for
    $k_\mathrm{max} = 2\times 10^{-9}\thinspace\text{eV}$ or
    $\Lambda_\mathrm{B}^{(\mathrm{min})} = 100\thinspace\text{m}$.
   (d) The behavior of the magnetic helicity density for the quenched $\Pi$
    with the parameters as in panel~(c) for shorter evolution time
    $t_0 < t < 5\times10^2\thinspace\text{yr}$.
    \label{fig:h}}
\end{figure*}

Along with growing magnetic fields, shown in Fig.~\ref{fig:B}, it is important to consider the evolution of the magnetic helicity density $h(t)=H(t)/V$ to illustrate its generation in a magnetar. In Fig.~\ref{fig:h} we demonstrate how the magnetic helicity grows in our model. We consider the cases of initially helical and nonhelical magnetic fields as well as quenched and unquenched parameter $\Pi$ in Eq.~\eqref{newPi} to compare our results with those in Ref.~\cite{DvoSem15b}. One can see in Figs.~\ref{2b} and~\ref{2d} that the difference in the evolution of initially helical and nonhelical magnetic fields is important only at early evolution times. Later this difference is washed out; cf. Figs.~\ref{2a} and~\ref{2b}. Therefore we extend the result of Ref.~\cite{DvoSem15b}, that the magnetic helicity can be generated in our model, to the case of  the quenched parameter $\Pi$ in Eq.~\eqref{newPi}.

In conclusion we mention that we have further developed the model, recently proposed in Refs.~\cite{DvoSem15a,DvoSem15b}, for the magnetic fields generation in magnetars. We have improved our approach pointing out that the magnetic field, growing owing to the instability caused by the parity violating $eN$ interaction, can take the energy mostly from thermal neutrons, as well as electrons and protons, which are present in the NS matter.

We have started with the evaluation of thermal corrections to the number densities and the energy densities of background fermions in NS. We have shown that, by cooling, these particles can pass their thermal energy to the magnetic field without violating the Pauli principle. Then, in the analogy with the solar dynamo, we have generalized the kinetic equations, derived in Ref.~\cite{DvoSem15b}, by quenching of the parameter $\Pi$; cf. Eq.~\eqref{newPi}. This procedure allowed us to treat background fermions as the large energy reservoir feeding the magnetic field. Moreover we have avoided the infinite growth of the magnetic field.

We have numerically solved the system of kinetic Eqs.~\eqref{general} with the modified parameter $\Pi$. For the initial conditions corresponding to a typical NS  ($n_{n,e}$ and $B_0$), we have obtained the growth of the seed magnetic field by three orders of magnitude to $B_\mathrm{sat} \approx 10^{15}\thinspace\text{G}$. Although this value of $B_\mathrm{sat}$ is smaller than that obtained in Refs.~\cite{DvoSem15a,DvoSem15b}, this $B_\mathrm{sat}$ is close to the magnetic field predicted in magnetars~\cite{Mer15}.

The time of the magnetic field growth to $B_\mathrm{sat}$ is $t_\mathrm{sat} = (10^4 - 10^5)\thinspace\text{yr}$ depending on the scale of the magnetic field. We have analyzed the two scales of the magnetic field in the range $\Lambda_\mathrm{B} = (10^2 - 10^3)\thinspace\text{m}$, i.e. we predict the generation of large scale magnetic fields. Comparing the obtained results for $t_\mathrm{sat}$ with the ages of magnetars~\cite{Mer15}, one concludes that our model is a quite plausible explanation of magnetic fields in magnetars.


We are thankful L.B.~Leinson and D.D.~Sokoloff for useful discussions. V.B.S. acknowledges G.~Sigl for comments on the subject. M.D. is grateful to the Competitiveness Improvement Program at the Tomsk State University and to RFBR (research project No.~15-02-00293) for partial support.


\begin{thebibliography}{99}

\bibitem{Maz79}
  E.~P.~Mazets, S.~V.~Golenetskij, and Y.~A.~Guryan,
  Soft gamma-ray bursts from the source B1900+14,
  Sov. Astr. Lett. \textbf{5}, 343 (1979).

\bibitem{FahGre81}
  G.~G.~Fahlman and P.~C.~Gregory,
  An X-ray pulsar in SNR G109.1-1.0, 
  Nature \textbf{293}, 202 (1981).

\bibitem{Fer15}
  L.~Ferrario, A.~Melatos, and J.~Zrake,
  Magnetic field generation in stars,
  Space Sci. Rev. (2015),
  doi:10.1007/s11214-015-0138-y;
  arXiv:1504.08074.

\bibitem{DvoSem15a}
  M.~Dvornikov and V.~B.~Semikoz,
  Magnetic field instability in a neutron star driven
  by the electroweak electron-nucleon interaction versus
  the chiral magnetic effect,
  Phys. Rev. D \textbf{91}, 061301 (2015);
  arXiv:1410.6676.

\bibitem{DvoSem15b}
  M.~Dvornikov and V.~B.~Semikoz,
  Generation of the magnetic helicity in a neutron star driven
  by the electroweak electron-nucleon interaction, 
   J. Cosmol. Astropart. Phys. 05 (2015) 032;
  arXiv:1503.04162.

\bibitem{BoyRucSha12}
  A.~Boyarsky, O.~Ruchayskiy, and M.~Shaposhnikov,
  Long-Range Magnetic Fields in the Ground State
  of the Standard Model Plasma,
  Phys. Rev. Lett. \textbf{109}, 111602 (2012);
  arXiv:1204.3604.


\bibitem{LanLif80}
  L.~D.~Landau and E.~M.~Lifshitz, 
  \textit{Statistical Physics: Part I}
  (Pergamon, Oxford, 1980), 3rd ed., pp.~168--171.

\bibitem{Nun97}
  H.~Nunokawa \textit{et al.},
  Neutrino conversions in a polarized medium,
  Nucl. Phys. B \textbf{501}, 17 (1997);
  hep-ph/9701420.

\bibitem{Cha14}
  P.~Charbonneau,
  Solar dynamo theory,
  Annu. Rev. Astron. Astrophys. \textbf{52}, 251 (2014).

\bibitem{Yak11}
  D.~G.~Yakovlev \textit{et al.},
  Cooling rates of neutron stars and the young neutron star
  in the Cassiopeia A supernova remnant,
  Mon. Not. R. Astron. Soc. \textbf{411}, 1977 (2011);
  arXiv:1010.1154.

\bibitem{Pet92}
  C.~J.~Pethick,
  Cooling of neutron stars,
  Rev. Mod. Phys. \textbf{64}, (1992) 1133.

\bibitem{Dav04}
  P.~A.~Davidson,
  \textit{Turbulence: An Introduction for Scientists and Engineers}
  (Oxford University Press, Oxford, 2004).

\bibitem{GraKapRed15}
  D.~Grabowska, D.~Kaplan, and S.~Reddy,
  The role of the electron mass in damping chiral magnetic instability
  in supernova and neutron stars,
  Phys. Rev. D \textbf{91}, 085035 (2015);
  arXiv:1409.3602.

\bibitem{SigLei15}
  G.~Sigl and N.~Leite,
  Chiral magnetic effect in protoneutron stars
  and magnetic field spectral evolution,
  submitted to J. Cosmol. Astropart. Phys.;
  arXiv:1507.04983.

\bibitem{MoiBis08}
  S.~G.~Moiseenko and G.~S.~Bisnovatyi-Kogan,
  Outflows from magnetorotational supernovae,
  Int. J. Mod. Phys. D \textbf{17}, 1411 (2008);
  arXiv:0801.2471.

\bibitem{Mer15}
  S.~Mereghetti, J.~A.~Pons, and A.~Melatos,
  Magnetars: Properties, origin and evolution,
  Space Sci. Rev. (2015),
  doi:10.1007/s11214-015-0146-y;
  arXiv:1503.06313.

\end{thebibliography}
\end{document}